\documentclass[12pt]{article}
\usepackage{graphicx}

\def\Title#1{\begin{center} {\Large #1 } \end{center}}
\def\Author#1{\begin{center}{ \sc #1} \end{center}}
\def\Address#1{\begin{center}{ \it #1} \end{center}}

\newenvironment{Abstract}{\begin{quotation}  }{\end{quotation}}

\begin{document}

\Title{Once More on Coulomb-Nuclear Interference}

\Author{ Vladimir A. Petrov } 
\Address{ Logunov Institute for High Energy Physics, NRC Kurchatov Inst., Protvino, RF}

\begin{center}
 E-mail: {Vladimir.Petrov@ihep.ru}
 \end{center} 

\begin{Abstract}
This is a critical reconsideration of the standard way of account for Coulomb-nuclear interference in the elastic scattering amplitude.
\end{Abstract}

\def\thefootnote{\fnsymbol{footnote}}
\setcounter{footnote}{0}

\section{Introduction}
As is well known, electromagnetic effects - soft photon radiation and Coulomb scattering - are an inseparable part of any strong interaction process with charged hadrons. Sometimes they hamper the observation of specific strong interaction phenomena but sometimes they are a unique source of information on important details of hadronic amplitudes.
In holographic terminology, the Coulomb interaction between colliding hadrons serves  an analogue of "mirror" giving the "reference wave" (the proton scattered via Coulomb exchange) whose interference with the "illuminating wave" (the proton scattered by strong forces) is to give a spatially complete image of an object
due to recording of the relative phase. Unfortunately, the interval of scattering angles where such interference is well seen is quite narrow but, nonetheless, analysis of the differential cross-section in this interval is capable to give us some very important information on the phase of the strong interaction amplitude\cite{Kru}.

The general form of the elastic hadron-hadron scattering(which is actually an inclusive process
with missing soft photons) differential cross-section is
\begin{equation}
\frac{d\sigma}{dt} = \frac{1}{16 \pi s^{2}}\mid T_{C+N} (s,t)\mid^{2}e^{-\Delta(t)}.
\end{equation}
where $ T_{C+N} (s,t) $ includes both strong interaction which in the absence of QED effects is $ T_{N} (s,t) $ and Coulomb exchanges while the damping factor $ \Delta(t) $ suppresses the cross-section due to the soft photon radiation. This damping factor is well studied (see, e.g.\cite{We}) while the way of accounting for the Coulomb contribution in $ T_{C+N} (s,t) $ "laisse \`{a} d\'{e}sirer".

Since the pioneering Bethe paper  \cite{Bethe} the extraction of the real part (or the phase) of the "ideal" scattering amplitude $  T_{N} (s,t)$ from the data (or, more often, a verification of its theoretically predicted value) was being based (see, e.g.\cite{We},\cite{Pr})  on the formula
\begin{equation}
T_{C+N} (s,t) = T_{C} e^{i\alpha\varphi(s,t)}+ T_{N} (s,t)
\end{equation}
where $ T_{C} $ is the lowest order Coulomb amplitude which for identical and point-like charges is well known: 
\begin{center}
$ T_{C} = 8\pi\alpha s /t$.
\end{center}
The Bethe ansatz (1) was being allegedly derived by several authors (see, e.g.\cite{Is}) but, to our mind, this was rather a set of plausible and phenomenologically attractive justifications than a mathematically consisted deduction from generally accepted premises. Some new steps free of these deficiencies were made in \cite{Ku}. Nonetheless, we have found necessary to again thoroughly derive the formulas for the total scattering amplitude with account of both strong and Coulomb interactions and we obtain some results which differ from those in \cite{Ku}. The first question we ask is: "Whether the Bethe ansatz holds formally, as a mathematically exact statement?"

\section{If the Bethe Anzats holds generally? }
For definiteness in what follows we will deal with the proton-proton scattering. Moreover, in this Section we consider protons as electrically point-like.Such a simplification would be justified if "the distance of closest approach" of the colliding protons were significantly larger than the proton proper size. This could be arranged at very small values of $ t $ , say $ -t = \mathcal{O} (10^{-6}GeV^{2}) $. We'll estimate the practical feasibility of such an opportunity in the last Section where we'll account for the proton form factor.

As to how exactly the strong and Coulomb interactions of protons are combined in the total scattering amplitude $T_{C+N} $ several options were in use starting from the naive addition of $  T_{N} $ and $ T_{C}$.
A more involved basic assumption is that in the eikonal representation (here $ b $ and $ q $ are 2D Euclidean vectors)
\begin{center}
$ T_{C+N} (s,t) = 2is\int d^{2}b e^{i{q}{b}}(1- e^{2i\delta_{C+N} (s,b)}) , t\approx -{q}^{2}, $
\end{center}
the eikonal function $ \delta_{C+N} (s,b) $ is additive w.r.t. strong ("nuclear", N) and electromagnetic ("Coulomb", C) interactions ("additivity of the strong and Coulomb potentials"):
\begin{equation}
\delta_{C+N} (s,b) = \delta_{C} (s,b)+\delta_{N} (s,b) .
\end{equation}
Such an additivity shown in Eq.(3) is not evident. In principle, one could add an "irreducible" term $ \sim \alpha $ which contains also strong interaction but is not a product of strong and e.-m. interaction terms and cannot be reproduced in the expansion of the exponential $ e^{2i\delta_{C+N} (s,b)} $ in a series in $\delta_{C+N}  $.  However, in this note we will not develop this subject leaving it for another occasion.

In Eq.(3) both eikonal functions $ \delta_{C} $ and $ \delta_{N} $ are defined by their "Born amplitudes". For the Coulomb phase we get
\begin{center}
$ \delta_{C} (s,{b}) = -\int \frac{d^{2}q}{(2\pi)^{2}} e^{-i{bq}} \frac{2\pi\alpha}{q^{2} + \lambda^{2}} = -\alpha K_{0}(\lambda \mid b\mid),  $
\end{center}
where $ \alpha $ is the fine structure constant, $ \lambda $ regularizes the Coulomb infrared singularity and $ K_{0}(z) $ is the Macdonald function of the zero order. Notice that when taking Fourier transform from $ {q}- $ to $ {b} $-spaces
we extend the maximum value of $ \mid {q} \mid$ to $ \infty $ in contrast to some authors who retain the maximum value as a heritage of the relation $ t = -2p^{2}(1-\cos \theta) $. We believe that at high energies and for the soft scattering such an account does not introduce changes of fundamental character but, instead,  allows a simple and convenient use of Fourier transformation to and fro.
Physical case of massless photons is retrieved at $ \lambda \rightarrow 0 $.
In this article we don't explicitly specify the strong interaction part, only tacitly use a rapid fall of $ T_{N} $ with growth of $ -t $.

Let us now put $|{q}|\neq 0  $. Then 
\begin{equation}
 T_{C+N} = -2is\Xi^{\alpha}({q})  + \int \frac{d{q}^{'}}{(2\pi)^{2}}\Xi^{\alpha}({q}-{q}^{'}) T_{N}(s,{t}^{'})  
\end{equation}
where
\begin{center}
$\Xi^{{\alpha}}({q}) = \int d^{2}{b} e^{i{qb}}\exp (- 2i\alpha K_{0}(\lambda \mid b\mid)).$
\end{center}

At $ \lambda\rightarrow 0 $ we use the approximation $  K_{0}(z)|_{z\rightarrow 0} \approx - \ln (z/2) - \gamma$ with $ \gamma = 0.5772... $ and the expression for $ \Xi^{{\alpha}}({q}) $
simplifies to 
\begin{center}
$  \Xi^{{\alpha}}({q}) = \int d^{2}{b} e^{i{qb}}(\frac{\bar{\lambda} \mid b \mid}{2})^{2i\alpha} = \lbrace\frac{\Gamma(1+i\alpha)}{\Gamma(1-i\alpha)}(\bar{\lambda}^{2}/q^{2})^{i\alpha}\rbrace \frac{4i\pi\alpha s}{t},  \bar{\lambda}\rightarrow 0 $
\end{center}
where $ \bar{\lambda} = \lambda exp(\gamma) \approx 1.78 \lambda $.
Making similar manipulations with the second term in $ T_{C+N} $ ( note that in this case the argument of $ \Xi^{{\alpha}} $ may assume zero value) we arrive at the following expression:
\begin{center}
$ T_{C+N}(s,t) = \lbrace\frac{\Gamma(1+i\alpha)}{\Gamma(1-i\alpha)}(\bar{\lambda}^{2}/q^{2})^{i\alpha}\rbrace [ \frac{8\pi\alpha s}{t} + \int\frac{d^2 q^{'}}{(2\pi)^2}C_{\alpha}({q} \vert{q}-{q}^{'})T_{N}(s,{t}^{'})] .$
\end{center}
The expression in braces is a pure phase, so we obtain for the modulus of the full amplitude the expression free of fictitious photon mass

\begin{equation}
\vert T_{C+N}(s,t)\vert = \vert\frac{8\pi\alpha s}{t} + \int\frac{d^2 q^{'}}{(2\pi)^2}C_{\alpha}({q}|{q}-{q}^{'})T_{N}(s,{t}^{'})\vert .
\end{equation}

Here the integral operator $ C_{\alpha} $  has the kernel
\begin{center}
$ C_{\alpha}({q}|{q}-{q}^{'})= -4i\pi \alpha |{q}-{q}^{'}|^{-2} [{q}^{2}/|{q}-{q}^{'}|^{2}]^{i\alpha}. $
\end{center}
Note that \cite{Gel'fand} 
\begin{center}
$ \lim\vert_{\alpha\rightarrow 0 }C_{\alpha} = (2\pi)^{2} \delta({q} - {q}^{'}). $
\end{center}

In terms of invariant variables we have
\begin{equation}
\vert T_{C+N}(s,t)\vert = \vert\frac{8\pi\alpha s}{t} -{i\alpha} \int dt^{'}\frac{\mid t \mid^{i\alpha}}{\vert t-t^{'}\vert^{1+i\alpha}}P_{i\alpha}(-\frac{t+t^{'}}{\vert t-t^{'}\vert}) T_{N}(s,t^{'})\vert .
\end{equation}
Here $ P_{\nu}(z)$ is the Legendre function of the first kind.

Expression (4) may be also presented in a bizarre but seemingly simple pseudo-differential form

\begin{equation}
\vert T_{C+N}(s,t)\vert = \vert\frac{8\pi\alpha s}{t} + q^{2i\alpha} ( - \nabla_{q}^{2})^{i\alpha}T_{N}(s,{t})\vert.
\end{equation}

Here  $\nabla_{q}^{2} $ is the Laplace operator in the 2D transverse momentum space.
The existence of the convolution 
\begin{center}
$  \int d^{2}q^{'} T_{N}({t}^{'}, s)\vert {q}-{q}^{'}\vert^{-2-2i\alpha}$
\end{center}
is ensured by  the fact that \cite{Gel'fand} 
\begin{center}
$ \int d^{2}q^{'}\vert {q}-{q}^{'}\vert^{-2-2i\alpha}= 0 .$
\end{center}

If the Bethe ansatz (1) were  true then we would have
\begin{equation}
q^{2i\alpha}( - \nabla_{q}^{2})^{i\alpha}T_{N}(s,{t})= \exp (-i\alpha \varphi (s,t)) T_{N}(s,{t}). 
\end{equation}
However, the simplest examples show that this is hardly the case. 

Let us take
a toy amplitude for $ T_{N} $: massive vector exchange
\begin{center}
$ T_{N} = 2g^{2}s/( M^{2}-t).$
\end{center}
We get in this case
\begin{center}

$ q^{2i\alpha}( - \nabla_{q}^{2})^{i\alpha}T_{N}(s,t)=\Phi_{i\alpha}T_{N}(s,t)$
\end{center}
where
\begin{center}
$\Phi_{i\alpha}=[(-2t/(M^{2}-t))^{i\alpha}\Gamma^{2}(1+i\alpha)P_{i\alpha}((M^{2}+t)/(M^{2}-t))].$
\end{center}
 
It is easy to verify that $ \vert\Phi_{i\alpha}\vert \neq 1  $ so $ \Phi_{i\alpha} $ cannot be of the form $ \exp (-i\alpha\varphi) $.

More realistic amplitude is, e.g., $ T_{N}(s,t)= is\sigma_{tot}\exp  (B(s)t/2) $ with $ B $ a "forward slope". Applying the operator $ q^{2i\alpha}( - \nabla_{q}^{2})^{i\alpha} $ we get the following factor $ \Phi_{i\alpha} $ :
\begin{center}
$ \Phi_{i\alpha} = (-2Bt)^{i\alpha}\Gamma (1+i\alpha)_{1}F_{1} (-i\alpha;1;-Bt/2) ,\ $
\end{center}
where $ _{1}F_{1} (\textit{a};\textit{b};z) $ is the confluent hypergeometric function.
We see again that the factor $ \Phi_{i\alpha} $ is not a pure phase.

Generally, taking into account that the operator $ (-\nabla_{q}^{2})^{i\alpha} $ is unitary it could seem plausible that the amplitude $ T_{N}(s,t) $ could be its eigenfunction with an eigenvalue $ \exp (-i\alpha \phi(s)) $ where $\phi(s)$ can't depend on $ t $. Now the Bethe phase would look as
\begin{center}
$ \varphi(s,t)= \ln \frac{s}{-t} +\bar{\varphi}(\frac{s}{M^{2}}) $ 
\end{center}
 with $ M $ a typical hadronic mass (e.g. the pion's). It is curious that the first term exactly coincides with an elegant expression suggested once by L. D. Soloviev \cite{Sol} in a fully relativistic context :
 \begin{center}
 $  \varphi_{Soloviev}(s,t)= \ln \frac{s}{-t} = 2\ln \frac{2}{\theta}$
 \end{center}
 with $ \theta $ the c.m.s scattering angle. Though afterwards this expression was found to be insufficient.
 
  Howbeit, in the "eigen-function" option the impact parameter amplitude would be of the form
  \begin{center}
   $ \tilde{T}_{N}(s,b)\sim \delta ( b^{2} - \frac{1}{M^{2}}\exp (-\bar{\varphi}(\frac{s}{M^{2}})) $
   \end{center}  which does not look very appealing.
   
Such are our arguments in favour of the statement that the Bethe ansatz (1) in its exact form does not take place.
\section{$ O(\alpha)$ approximation  }
In principle, formulas like Eq.(5) could be used directly. However, practically it is not easy to deal with functions like $ P_{i\alpha} (z) $.
Moreover, since the pure Coulomb term in Eq.(5) is of the first  order in $ \alpha $ it seems justified and natural to see what happens with the mixed term when we expand it in powers of $ \alpha $.

However, when trying to do that we immediately encounter an obstacle: a straightforward expansion of the kernel $ C_{\alpha}({q}|{q}-{q}^{'}) $ in powers of $ \alpha $ under the integral leads to non-integrable expressions of the type $ 1/\mid{q}\mid^{2} $ which are actually generalized function (distributions):
 \begin{center}
 $C_{\alpha}({q}|{q}-{q}^{'}) = (2\pi)^{2} \delta ({q}-{q}^{'}) -4i\pi\alpha \mid {q}-{q}^{'}\mid ^{-2} +  ...$
 \end{center}
Thus, the generalized function  $ 1/\mid{q}\mid^{2} $ is well defined on the subspace of test functions which disappear at $ {q} =0 $. To arrange and use such a property we note that at  real $ q $ the scattering amplitude ${T}_{N}(s,t)$ can be locally considered as a "test function" due to its belonging to the class $ C^{\infty}(R^{2}) $ and rapid decay at large $ \mid q\mid $. We also can take advantage from the identity 
\begin{eqnarray}
  \int \frac{d^{2}q^{'}}{(2\pi)^{2}}C_{\alpha}({q}|{q}-{q}^{'})\theta(\mid{q}\mid - \mid {q} - {q }^{'}\mid) = 1 
\end{eqnarray}
which is easy to check. In fact,
\begin{center}
$ \int \frac{d^{2}q^{'}}{(2\pi)^{2}}C_{\alpha}({q}|{q}-{q}^{'})\theta(\mid{q}\mid - \mid {q} - {q }^{'}\mid) =
-2i\alpha\int_{0}^{\mid q \mid} \frac{dq^{'}}{q^{'}}(\frac{q}{q^{'}})^{2i\alpha} $
\end{center}
\begin{center}
$  = -2i\alpha \int_{0}^{1} d\xi \xi^{-1-i\alpha}=\xi^{-i\alpha}\mid_{0}^{1}=1-\lim _{\xi\rightarrow 0} e^{i\alpha \ln \frac{1}{\xi}}. $
\end{center}
Taking into account that $ e^{ix\tau}\rightarrow 0 , \tau\rightarrow\infty, \forall x\neq0 $ \cite{Vla} we arrive at Eq.(5).
This identity enables us to put Eq.(4) in the form:
\begin{flushleft}
$\vert T_{C+N}(s,t)\vert = $
\end{flushleft}
\begin{center}
$=\vert\frac{8\pi\alpha s}{t} + T_{N}(s,t) + 
\int\frac{d^2 q^{'}}{(2\pi)^2}C_{\alpha}({q}|{q}-{q}^{'})[T_{N}(s,{t}^{'})- T_{N}(s,{t})\theta(\mid{q}\mid - \mid {q} - {q }^{'}\mid)]\vert.$
\end{center}
Now the expansion in $ \alpha $ in the integrand is harmless and we obtain
\begin{center}
$ \vert T_{C+N}(s,t)\vert = \vert\frac{8\pi\alpha s}{t} + T_{N}(s,t) -4i\pi\alpha 
\int\frac{d^2 q^{'}}{(2\pi)^2}\frac{[T_{N}(s,{t}^{'})- T_{N}(s,{t})\theta(\mid{q}\mid - \mid {q} -{q }^{'}\mid)]}{\mid {q}-{q}^{'}\mid^{2}}\vert + \textit{O}(\alpha^{2}) $
\end{center}
In terms of invariant transfers we get
\begin{equation}
\mid T_{C+N}(s,t)\vert = 
\vert\frac{8\pi\alpha s}{t} + T_{N}(s,t) -i\alpha 
\int_{-\infty}^{0} {dt{'}}\frac{1}{\mid t-t^{'}\mid }[T_{N}(s,t^{'}) - T_{N}(s,t)\kappa (t,t^{'})] 
\end{equation}
where
\begin{center}
$\kappa (t,t^{'}) = \theta (t^{'}-4t) \frac{2}{\pi} \arctan [\tan (\chi/2)(\sqrt{-t}+\sqrt{-t^{'}})/\mid\sqrt{-t}-\sqrt{-t^{'}}\mid ] , \cos \chi = \frac{1}{2}\sqrt{t^{'}/t}.$
\end{center}
Note that $ \kappa (t,t) = 1 $.
In Ref.\cite{Ku} the expression for $ \vert T_{C+N}(s,t)\vert $ in the case of point like charges and in the same approximation in $ \alpha $ looks a little differently
\begin{equation}
\vert T_{C+N}(s,t)\vert = 
\vert\frac{8\pi\alpha s}{t} + T_{N}(s,t) -i\alpha 
\int^{0} {dt{'}}\frac{1}{\mid t-t^{'}\mid }[T_{N}(s,t^{'}) - T_{N}(s,t)]\mid 
\end{equation}
We again draw attention of the reader to the fact that when integrating in $ {q} $ or $ t $ we consider all the space (${q}\in R^{2}, -t \in(0,+\infty )$) including some unphysical values (e.g., $-t\in (s-4m^{2},+\infty )$. We believe that due to the soft character of strong interactions high space-like momenta do not play significant role. In our reasonings above we never face UV divergencies. This enormously simplifies the use of the 2D Fourier transformations.
On the contrary, in \cite{Ku} the high $ - t^{'} $ divergence in Eq.(10) is cured by retaining the kinematic upper limit for $ \mid t^{'}\mid \sim s $ as was mentioned above. Such a strong influence of  this "ultraviolet" divergence is , in our opinion, alien to a typically soft framework.
\section{Account of the form factors }
 The formula obtained in the preceding Section could be only used if the average distance between colliding protons would significantly exceed their "proper sizes" which are naturally  identified with their average valence core radii.
 The latter can be estimated (in 2D projection) as (see, e.g.\cite{Pe}) 
 \begin{center}
 $  \langle b^{2}\rangle_{proton} \approx(0.66 fm)^{2} = 11.20 GeV^{-2}. $
 \end{center}
 The TOTEM Collaboration \cite{TO} gives for the average distance $\langle b^{2}\rangle $ between the centres of the colliding protons at $ \sqrt{s}  = 7 TeV$ the value 
 $\langle b^{2}\rangle\vert_{TOTEM} \approx (1.25 fm)^{2} $. We see that we are still very far from having right to neglect the proper sizes of the colliding protons.This means that we have to modify our expressions with including the proton electric form factors $ F(q^{2}) $ in the Coulomb eikonal phase. Actually these form factors can be identified with "effective form factors" as was introduced in \cite{proch}. We have now
 \begin{equation}
 \delta_{C} (s,{b}) = -\int \frac{d^{2}q}{(2\pi)^{2}} e^{-i{bq}} \frac{2\pi\alpha}{q^{2} + \lambda^{2}} F^{2}(q^{2}). 
 \end{equation}
 Eq.(4) for the total amplitude $ T_{C+N} $ remains valid in its general form but now in function $ \Xi^{\alpha} $ we keep $ \delta_{C} $ as in Eq.(12).
 To eliminate the dependence on the fictitious photon mass $ \lambda $ we use the old trick (see, e.g.\cite{La} ).  Let us take again  $ q\neq 0 $. This allows us to factor out the common factor $ exp (2i \delta_{C}(b=0))$ and we are being left with 
\begin{equation}
\mid T_{C+N}(s,t) \mid = \mid -2is\hat{\Xi}^{\alpha}(q)+ \int\frac{d^{2}q^{'}}{(2\pi)^{2}}\hat{\Xi}^{\alpha}(q-q^{'})T_{N}(s,t^{'}) \mid
\end{equation}
where
\begin{center}
$\hat{\Xi}^{\alpha}(q) = \int d^{2}b e^{iqb} e^{2i\hat{\delta}_{C}(b)}, 
\;\hat{\delta}_{C}(b)= -\alpha \int_{0}^{\infty}\frac{d\mid q\mid}{\mid q\mid} F^{2}(q^{2})[J_{0}(\mid q\mid \mid b\mid)-1].$ 
\end{center}
In these formulas high-$q$ convergence is provided by the fast decrease of the form factors while the subtraction helps to cure IR and to send $ \lambda $ to 0.
The function $ \hat{\Xi}^{\alpha}(q) $ cannot be represented in a closed analytic form for existing model forms of the form factor. 
It is easier to deal with the lowest order in the fine structure constant. However, in the same way as it took place in the point-like case the straightforward expansion in $ \alpha $ will lead to functions non-integrable at $ q=0 $. To circumvent this problem we notice a property:
\begin{equation}
\int \frac{d^{2}q}{(2\pi)^{2}} \:\hat{\Xi^{\alpha}}(q) = 1
\end{equation}
which is easy to verify.
This enables us to make the following identical transformation of Eq.(13):
\begin{center}
$\mid T_{C+N} (s,t) \mid = \mid -2is\hat{\Xi}^{\alpha}(q)+ T_{N}(s,t) - T_{N}(s,t)\cdot\int\frac{d^{2}q^{'}}{(2\pi)^{2}}\hat{{\Xi}}^{\alpha}(q^{'})$
\end{center}
\begin{equation}
+\int\frac{d^{2}q^{'}}{(2\pi)^{2}}\hat{\Xi}^{\alpha}(q-q^{'})T_{N}(s,t^{'}) \mid 
\end{equation}
and then to come to the form which allows the regular expansion in $ \alpha $ under the integral sign:
\begin{equation}
\mid T_{C+N}(s, t) \mid = \mid -2is\hat{\Xi}^{\alpha}(q)+ T_{N}(s,t)+\int\frac{d^{2}q^{'}}{(2\pi)^{2}}\hat{\Xi}^{\alpha}(q-q^{'})[T_{N}(s,t^{'})- T_{N}(s,t)]\mid .
\end{equation}
So the first terms up to $ \mathcal{O}(\alpha^{2})$ are:
\begin{equation}
\mid T_{C+N}(s,t) \mid = \mid T_{N}(s,t) + (8\pi s \alpha)/t 
+ \frac{i\alpha}{2\pi}\int dt^{'}[T_{N}(s,t^{'})- T_{N}(s,t)]I(t,t^{'})\mid
\end{equation}
where
\begin{center}
$ I(t,t^{'}) = \int_{0}^{2\pi} d\psi \frac{F^{2}(t+t^{'}-2\sqrt{tt^{'}}\cos \psi) }{(t+t^{'}-2\sqrt{tt^{'}}\cos \psi)} $
\end{center}
or in a more explicit form:
\begin{center}
$ I(t,t^{'})= 2\int \frac{d\kappa^{2}}{\kappa^{2}}F^{2}(\kappa^{2})(\kappa^{2}_{+}(t,t^{'})-\kappa^{2})_{+}^{-1/2}(\kappa^{2}- \kappa^{2}_{-}(t,t^{'}))^{-1/2}_{+}  $
\end{center}
where
\begin{center}
$ x^{\nu}_{+}  = x^{\nu} \vartheta (x),\kappa^{2}_{\pm}(t,t^{'})=(\sqrt{-t}\pm\sqrt{-t^{'}})^{2} .$
\end{center}

 For the same quantity, $ \mid T_{C+N} \mid $ in the leading order in $ \alpha $, we find in \cite{Ku} (in our notations) the expression:
 \begin{center}
$ \mid T_{C+N} \mid = \mid  T_{N}(s,t)+(8\pi s \alpha)/t 
 +\frac{i\alpha}{2\pi}\int dt^{'}[T_{N}(s,t^{'})- T_{N}(s,t)]I(t,t^{'})$
  \end{center}
 \begin{equation}
  -i\alpha T_{N}(s,t)\int dt^{'}\ln (\frac{t^{'}}{t})\frac{d}{dt^{'}}F^{2}(t^{'}))\mid
  \end{equation} 
which contains an extra term absent from Eq.(17).
A similar term was earlier obtained by R. Cahn (see Eq.(30) in the second item in Ref. \cite{Is}).
The root of this discrepancy lies in a wrong expression of the Coulomb term "dressed" with multiple exchanges in the presence of non trivial form factors (cf. Eq.(22) in Cahn's paper).

\section{Conclusion}
 Above we have presented arguments against a general validity of the "Bethe ansatz" (2) for the Coulomb-nuclear amplitude $ T_{C+N} $. 
 We also gave some new derivations of the form of $ \mid T_{C+N}\mid $, both for the point-like and distributed charges, with results which differ significantly from those given by other authors.
As indicated in \cite{TOT} exactly Eq.(18) was used for extraction of the sensationally small ($ 0.09 \div 0.10 $) value of the parameter $\rho = \arctan Arg (T_{N}(s,t=0)) $ at $ \sqrt{s}=13 TeV $.
We do not see a flaw in derivation of our Eq.(17) so it would be very interesting to know in which way its use can influence the value of $ \rho $ when being extracted from the data.

 I am grateful to Jan Ka\v{s}par and Anatoliy Samokhin for useful discussions. 
   
This work is supported by the RFBR Grant 17-02-00120.


\begin{thebibliography}{99}

\bibitem {Kru}
V. Kundr\'{a}t and M. Lokaj\'{i}\v{c}ek and D. Krupa, Czech.J.Phys.\textbf{B39},
1245 (1989).
\bibitem {We}
G.B. West and D. R. Yennie, Phys. Rev. \textbf{172}, 1413(1968).
\bibitem{Bethe}
H. Bethe, Ann. Phys. 3, 190(1958){\bf 13}.
\bibitem{Pr}
O. V. Selyugin, Mod.Phys.Lett. \textbf{A11},2317 (1996);

V. A. Petrov, A. V. Prokudin and E. Predazzi, Eur.Phys.J.\textbf{ C 28}, 525 (2003).
\bibitem{Is}
M. M. Islam, Phys. Rev.\textbf{162},1426(1967);

R. Cahn, Z. Phys. \textbf{C15}, 253 (1982).
                       
\bibitem{Ku}
V. Kundr\'{a}t and M. Lokaj\'{i}\v{c}ek, Z. Phys. \textbf{C63}, 619(1994).

\bibitem{Gel'fand} I.M. Gel'fand and G. E. Shilov. Generalized Functions.

AMS Chelsea Publishing.Volume1; 1964. 
\bibitem{Vla} V.S. Vladimirov. Equations of Mathematical Physics. 

"Mir" Publishers, Moscow,1984.
\bibitem{Sol}
L.D.Soloviev, JETP \textbf{22},205(1966).
\bibitem{Pe}
V.A. Petrov, arXiv:1711.00226.
\bibitem{TO}
TOTEM Collaboration, Eur.Phys. J, \textbf{C76}, 661 (2016).
\bibitem{proch}
J. Proch\'{a}zka and V.Kundr\'{a}t, arXiv:1606.09479
\bibitem{La}
L. D. Landau and E. M. Lifshits. Quantum Mechanics. \S136.

2nd Ed. Pergamon Press.

\bibitem{TOT}
G. Antchev et al. TOTEM Collaboration. Preprint CERN-EP-2017-335.
\end{thebibliography}
\end{document}